\documentclass[12pt]{iopart}

\usepackage{iopams}
\usepackage[dvips]{graphicx}
\usepackage{epsfig}

\begin{document}

\title[Scaling  of   deterministic  walks  in   complex  environments]
{L\'evy-like behaviour  in deterministic models  of intelligent agents
exploring heterogeneous environments}

\author{D Boyer$^{1,2}$, O Miramontes$^{1,2}$ and H Larralde$^3$}

\address{$^1$ Instituto de F\'\i sica, Universidad Nacional Aut\'onoma
de M\'exico, Apartado Postal 20-364, 01000 M\'exico D.F., M\'exico}

\address{$^2$ C3 - Centro de Ciencias de la Complejidad,
Universidad Nacional Aut\'onoma de M\'exico, Cd. Universitaria,
Circuito Escolar, M\'exico 04510 D.F., M\'exico}

\address{$^3$  Instituto de  Ciencias F\'isicas,  Universidad Nacional
Aut\'onoma  de  M\'exico,  Apartado  Postal  48-3,  Cuernavaca,  62251
Morelos, M\'exico}

\ead{\mailto{boyer@fisica.unam.mx},    \mailto{octavio@fisica.unam.mx},
\mailto{hernan@ce.fis.unam.mx}}

\begin{abstract}
Many  studies  on  animal  and  human  movement  patterns  report  the
existence  of scaling  laws  and power-law  distributions.  Whereas  a
number  of   random  walk  models   have  been  proposed   to  explain
observations, in  many situations individuals actually  rely on mental
maps  to explore  strongly  heterogeneous environments. 
In this work we study a
model of  a deterministic walker, visiting  sites randomly distributed
on the plane and with  varying weight or attractiveness. At each step,
the walker  minimizes a function that  depends on the  distance to the
next unvisited target (cost) and  on the weight of that target (gain).
If  the   target  weight   distribution  is  a   power-law,  $p(k)\sim
k^{-\beta}$,  in some  range  of the  exponent  $\beta$, the  foraging
medium induces  movements that are  similar to L\'evy flights  and are
characterized by  non-trivial exponents. We explore  variations of the
choice rule  in order to  test the robustness  of the model  and argue
that the  addition of noise  has a limited  impact on the  dynamics in
strongly disordered media.
\end{abstract}

\pacs{05.40.Fb,87.23.-n,75.10.Nr}
\submitto{\JPA}

\section{Introduction}

Deterministic   walkers,  as   opposed  to   random   ones,  follow
non-stochastic rules of  motion. They can be used  as an approximation
to describe mobile agents that  {\it (i)} have some information on the
medium  they  explore  (composed,   for  instance,  of  food  patches,
facilities or  cities, depending on  the context); and {\it  (ii)} use
this information  to optimize their gain  (such as food  intake or the
possibility  of realizing a  particular activity).   If the  medium is
disordered and  heterogeneous, deterministic rules  can produce fairly
erratic  and   complex  trajectories,  susceptible   of  being  studied
statistically as random walks.

Unlike  Brownian particles, humans  and other  animals keep  memory of
their  past activity.  For  example, a  class  of deterministic  walks
assumes that  each decision depends  on the previous steps  the walker
has performed. This  kind of processes can be applied  to a wide range
of  hard-to-solve problems  such  as the  travelling  salesman or  the
travelling  tourist \cite{lima2001,stanley2001}.   Yet,  despite  their
potential importance in a variety of problems in physics, mathematics,
computer   science,   economics  and   biology,   the  properties   of
deterministic walks remain largely unexplored. Deterministic diffusion
has  been investigated  in simple  random  media \cite{bunimovich2004,
tercariol2005}.   Similar   processes  can   be   used  to   implement
classification schemes  \cite{kinouchi2002,Backes2006} and for pattern
recognition \cite{Campiteli2006}. In  biology, though not widely used,
deterministic walks offer an appealing framework to model the movement
of foraging  animals. Indeed,  many organisms rely  on mental  maps to
navigate  their  environment in  a  non-random way  \cite{garber1989}.
Primates and  other animals dispose of  sophisticated cognitive skills
\cite{griffiths99}, such  as the ability  to remember autobiographical
events  that happened at  some particular  places and  times (episodic
memory).

The presence of scaling laws and power-law distributions in animal
movement patterns is an ecological problem that has attracted
considerable attention in the last decade \cite{visreview}. Empirical
evidence of L\'evy-walks has been reported, among others, for
micro-organisms \cite{bartumeus}, bumblebees \cite{vis99}, spider
monkeys \cite{gabriel}, marine predators \cite{sims}. Similar patterns
have also been observed in humans, as shown by studies on
hunter-gatherers \cite{brown}, bank notes \cite{geisel} or fishermen
boats \cite{peru}.  Human individuals also display multiple scale
displacements well described by truncated power-laws \cite{barabasi}.
Most of the theoretical interpretations proposed to explain
observations rely on the random walk hypothesis or some variant
thereof: for example, the diffusion of bank notes is
actually well described by a continuous time random walk model
\cite{geisel}.

A common hypothesis is that the ubiquitous presence of
widely fluctuating characteristic scales in animal movement could
optimize the success of search processes of randomly distributed
resources \cite{vis99}.  According to this perspective, animals would
execute a particular optimal foraging strategy, usually a Markovian
stochastic process (see \cite{vis99,benichou,klafterpnas} for
examples), to find preys that are uniformly distributed and only
detectable at short distance. These strategies are susceptible of
being learned along an individual life experience \cite{sims}, or
transmitted through generations as an evolutionary trait
\cite{bartumeusfractal}.

An alternative hypothesis, namely, that scale-free (or more
generally, multiple scale) movement patterns may be an outcome of the
direct interaction of the forager with a complex environment, has been
much less explored quantitatively.  In this approach, very simple
movement rules, sometimes involving memory, may produce complex
searching patterns.  For instance, a simple response to
passive scent concentrations in turbulent structures can account for
scale-free trajectories of flying insects \cite{reynoldspre2005}.
L\'evy patterns similar to those observed in jackals can also arise
from avoidance of conspecific odour trails, as modelled by systems of
self-avoiding walks \cite{reynoldsepl}.  L\'evy-like distributions of
the distances between detected preys can also emerge in the
trajectories of a predator following, deterministically, chemotaxis
gradients produced by immobile and randomly located preys
\cite{reynoldspre2008}.

In this article, we study a model describing the movement of a walker
that forages in a strongly heterogeneous environment using mental
maps.  It is assumed that the forager has a perfect knowledge of the
available resources (for instance, fruiting trees) and that it chooses
deterministically at each step the \lq\lq best" (in terms of an
efficiency function) unvisited food patch in the whole system.
Therefore, decisions are not taken on the basis of the knowledge of
the immediate surroundings only.  We originally proposed this model in
the context of a study on the ranging patterns of spider monkeys in
tropical forests \cite{boyer2006,boyer2005}, but we expect it to have
a much wider application range. In that work, we showed that
scale-free distributions of step lengths can emerge if the resources
are sufficiently diverse in size. Here we give a detailed
analysis of the model, we show that the behavior of the
walker is robust to the addition of noise and we study how the
results are affected by changes in the deterministic decision
rules.

\section{The model}

Consider a two-dimensional square domain of unit area, randomly and
uniformly filled with $N$ point-like targets of fixed positions. The
targets model, for instance, fruiting trees in an ecological context,
or cities in a social one.  The targets may also represent network
communities, nodes for information retrieval or markets for salesmen.
Each target $i$ is characterized by, say, a food content or a size,
$k_i$, which are independent random integers drawn from a distribution
$p(k)$. (Integers are considered here for computational convenience,
without restricting generality.)  The distributions of tree sizes in
tropical and template forests are very broad and in many cases well
described by inverse power-laws with exponents ranging from 1.5 to 4
\cite{enquist,niklas}. Across species, the size of a tree is roughly
proportional to its fruit mass \cite{snook}.  We will therefore
consider target size distributions of power-law form:
\begin{equation}\label{pk}
p(k)=Ck^{-\beta},\quad{\rm with}\ k=1,2,...,k_{max};\ \beta>1,
\end{equation}
$C$ being the normalization constant;  and $k_{max}$ is a cutoff size,
that can be taken infinite (see section \ref{1d}) but will be fixed to
$10^3$ in  the simulations below, except when  otherwise indicated. In
the  social context, (\ref{pk})  with $\beta=2$  is equivalent  to
assuming that  city sizes  are distributed according  to a  Zipf's law
\cite{newman}.

In this domain, a forager (or traveler) is initially located on a
target.  We assume that the forager has a perfect knowledge of the
size and position of every target in the system. In a human
context, when the travelers are, say, salesmen or tourists this
assumption is natural. On the other hand, in a
biological context, this assumption is based on observations that
many animals (primates \cite{garber1989,griffiths99}, birds
\cite{gould}) use local and global cues during searching. Field
studies show that Clark's nutcracker (among other species of small
birds) can store 22,000 to 33,000 nuts in up to 2700 locations over
an area larger than 100 square miles and remember where almost
$70\%$ of them where placed \cite{lanner}. One could also relax the
perfect knowledge hypothesis by assuming that the forager knows a
random subset of $N_s$ targets, or the $N_s$ largest targets ($1\ll
N_s<N$). These known targets would still be described by the
distribution (\ref{pk}) with exponent $-\beta$.

At  each time step ($t\rightarrow t+1$) the following rules are iterated:

{\it (i)} the walker located at  target $i$ will next visit the target
$j$  such that  the  step  efficiency $E_{ij}$  is  maximum among  all
allowed $j\ne i$ in the system,

{\it  (ii)}  previously  visited  targets  are  not  revisited  (which
corresponds to destructive foraging in the ecological context).

Presently, we will consider two forms for the efficiency
function $E_{ij}$:
\begin{eqnarray}
E_{ij}&=&k_j/l_{ij}\quad({\rm\lq\lq   quotient"\   rule})\label{kod}\\
E_{ij}&=&k_j-l_{ij}/l_0\quad({\rm\lq\lq difference"\ rule}),\label{kmd}
\end{eqnarray}
where  $k_j$  is  the  size  of  target  $j$,  $l_{ij}$  the  distance
separating  targets  $i$ and  $j$,  and  $l_0=1/\sqrt{N}$ the  typical
distance  between  nearest neighbour  targets.  Hence,  in both  rules
(\ref{kod})-(\ref{kmd}),  the  forager  tends  to  visit  the  largest
possible target  accessible in a  minimum travelled distance  from its
current position,  ignoring targets that  are either too small  or too
far  away. Rule  (\ref{kmd}), which  has  not been  studied before  in
\cite{boyer2005},  represents the  net  energy balance  of a  sojourn,
assuming that  the energy expenditure is proportional  to the distance
travelled. Note that in this case we do not impose the condition $E_{ij}>0$:
some steps may be such that energy expenditure exceeds energy gain 
(see section \ref{secbalance} below).

\section{Step lengths statistics}

In \cite{boyer2005}, a mean field, single-step solution of the
model for the quotient rule (\ref{kod}) was obtained in any spatial
dimension $d$.  There, it was shown that the heterogeneity
of the medium can induce power-law distributions $P(l)$ of sojourn
lengths (distances between the origin and the best target in that
case): In $d=2$, if $\beta\ge\beta_c=3$ in the target size
distribution (\ref{pk}), then
\begin{equation}
P(l)\propto l^{-\alpha_{MF}},\ {\rm with}\ \alpha_{MF}=\beta-2.
\end{equation}
Therefore   displacements    can   be   considered    as   L\'evy-like
($1<\alpha<3$) if $3<\beta<5$.  If $\beta<3$ and $k_{max}=\infty$, the
density of very big targets in the medium is so high that $P(l)$ is no
longer scale-free  and involves  a characteristic size  diverging with
$N$ \cite{boyer2005}.

Although the  mean-field description  captures the gross  behaviour of
the model,  it fails to  explain the numerical  results quantitatively
(see below).  By imposing the condition that the walker cannot return
to a previously visited  site, memory effects produce non-trivial long
range  correlations   between  steps  as  the   process  develops.  In
\cite{boyer2006},   $2d$  simulations   with  the   quotient  rule
(\ref{kod}) indeed exhibit scaling-laws, but different exponents are 
found:
\begin{equation}\label{scalnum}
P(l)\propto            l^{-\alpha},\            {\rm            with}\
\alpha\simeq\beta-1(\neq\alpha_{MF}),\ {\rm for}\ 3\le\beta\le 4.
\end{equation}

\begin{figure}
\hspace{2.7cm}                   \epsfig{figure=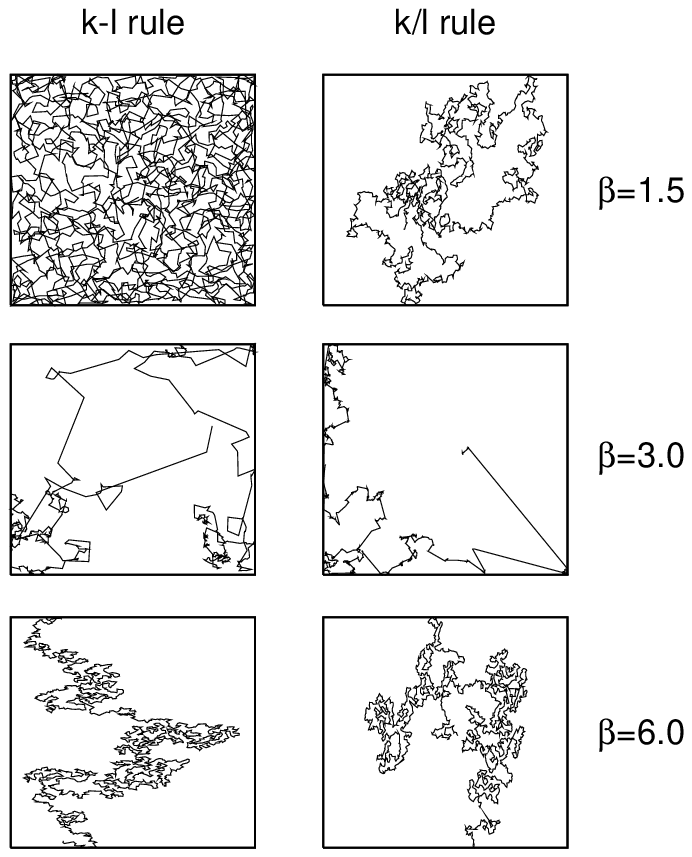,width=2.6in}
\epsfig{figure=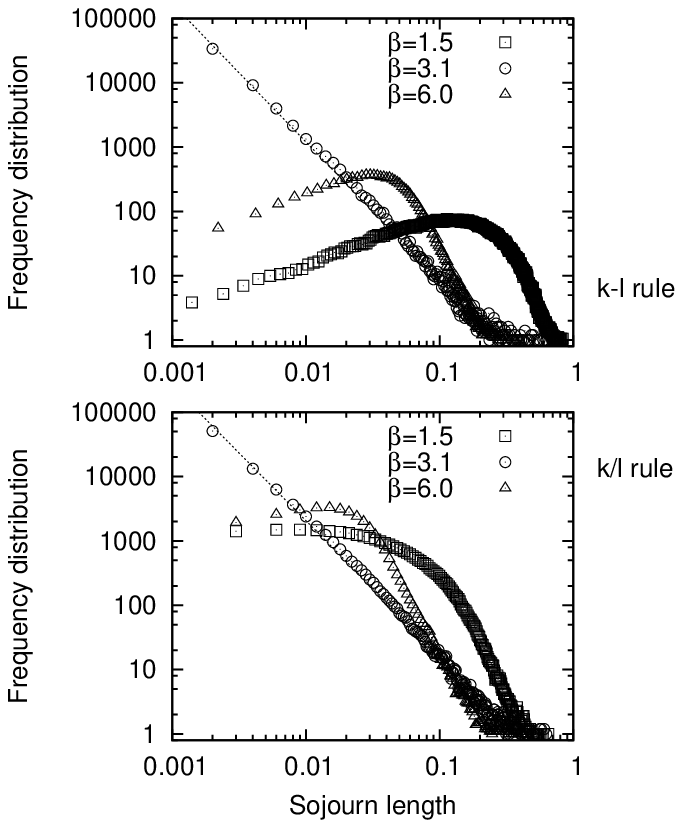,width=2.6in}
\caption{{\bf Left panels:} Examples of deterministic trajectories for
3  values  of  the  target  size  distribution  exponent:  $\beta=1.5$
(superabundant   medium),  $3$  (diverse   medium)  and   $6$  (homogeneous
medium). When $\beta=3$, the
sojourn length fluctuates widely in  both cases, similarly to a random
L\'evy  flight.  ($N=10^6$,  number of  steps per  walk$=10^5$.)  {\bf
Right  panels:}  Step  length  distributions.  Near-perfect  power-law
distributions,  $P(l)\sim  l^{-\alpha}$,  are  obtained in  the  range
$3\le\beta\le 4$. The value of  the scaling exponent at $\beta=3.1$ is
$\alpha\simeq2.1$ while the value at $\beta=4.0$ is $\alpha\simeq3.0$}
\label{trajectory}
\end{figure}

We discuss in more detail the numerical results below.

Several trajectories are represented in figure \ref{trajectory}-left,
with the corresponding step length distribution in figure
\ref{trajectory}-right.  With the \lq\lq difference" rule (\ref{kmd}),
three qualitative regimes can be identified as $\beta$ is varied.
In the first one
(superabundant regime), corresponding to $\beta<3$, the trajectories
are dominated by large sojourn lengths. 
If $\beta$ is small enough, there are many attractive
(large) targets, so that the forager often travels large distances to
reach them.  Very large targets are relatively distant from each
other, but the large values of their size $k$ make these long sojourns
\lq\lq affordable".  The situation changes dramatically in the
interval $3.1<\beta<4.0$ (diverse regime), where very big targets
exist but are too scarce to produce frequent long travels. As a
result, there are a few large sojourn lengths alternated with many
shorter ones at all scales.  In this regime, the distribution of step
lengths is a power-law with an exponent $\alpha$ approximately given
by relation (\ref{scalnum}). Trajectories in this regime closely
resemble L\'evy flights.  Finally, when $\beta>4$ (homogeneous
regime), target size fluctuations are too small and the forager
performs mostly short steps to nearby targets, similar to a (self
avoiding) random walker.

The three  regimes discussed above,  in particular the  scaling regime
with $\alpha\simeq\beta-1$, see (\ref{scalnum}), are also observed
when the  walker follows the  \lq\lq quotient" rule (\ref{kod}).  
In  the superabundant
regime, nevertheless,  the steps are  on average shorter than  for the
difference rule (see figure \ref{trajectory}-right).

\begin{figure}
\centering \includegraphics[width=0.5\textwidth]{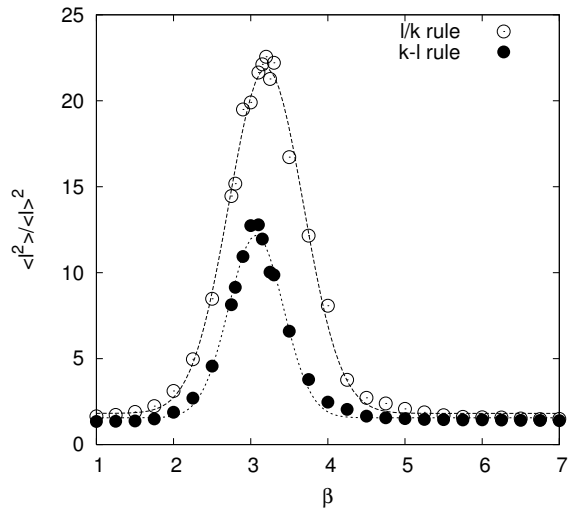}
\caption{Length   fluctuation  ratio,  $\langle   l^2\rangle/  \langle
l\rangle^2$, as a function of  the resource exponent $\beta$.  In both
the   $k-l/l_0$   and  $k/l$   rules,   this   quantity   is  maximum   at
$\beta\approx3.1$.   In all cases,  $N=10^6$ and  $k_{max}=10^3$. Data
was  obtained from averaging  over 10  different disordered  media. In
each realization, a  walk starts near the center  of the square domain
and visits $10^5$ targets, a number still $\ll N$.}
\label{figfluctu}
\end{figure}

\begin{figure}
\centering \includegraphics[width=0.5\textwidth]{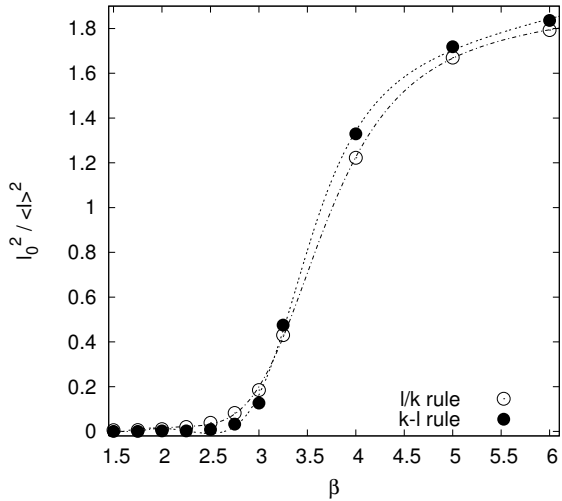}
\caption{ ${l_0}^2  / \langle  l \rangle^2$ as  a function  of $\beta$
showing diverging steps below  $\beta_c$. $k_{max}=10^6$,
other system settings as in figure \ref{figfluctu}.  }
\label{figl2}
\end{figure}

Further insight  into the statistical  properties of the walks  can be
gained  by studying step  length fluctuations.  Figure \ref{figfluctu}
displays  the  fluctuation   ratio,  defined  as  $\langle  l^2\rangle
/\langle   l\rangle^2$,   versus   the  resource   exponent   $\beta$.
Fluctuations  are  maximum   at  $\beta_c\approx3$  (independently  of
$k_{max}$) for both decision rules, confirming the results obtained in
\cite{boyer2006} for  rule (\ref{kod}) only. Therefore,  a medium with
$\beta=\beta_c$ produces,  in some  sense, the most  scale-free walks,
that  are  characterized  by   $\alpha\simeq2$.  As  shown  in  figure
\ref{figl2}, $\beta_c$ also corresponds  approximately to the onset of
emergence of very large steps: $(l_0/\langle l\rangle)^2$ becomes very
small below $\beta_c$.

\section{Sizes of visited sites and non-stationarity}

\begin{figure}
\centering \includegraphics[width=0.5\textwidth]{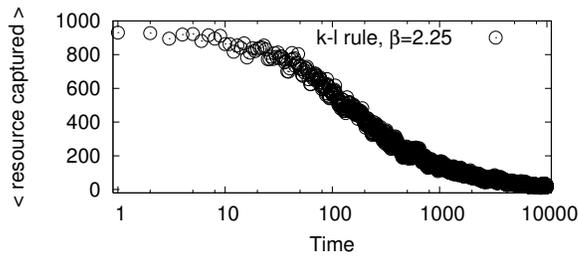}
\caption{Temporal evolution  of the average  size of the  $i^{\rm th}$
visited  target, in the  $k-l$ rule  and with  $\beta=2.25$, $N=10^6$,
$k_{max}=10^3$. Averages are taken over 10 independent runs.}
\label{figsizevisited}
\end{figure}

We now  analyze the size of  visited targets, which is  related to the
intake   of   the  forager   in   the   ecological  context.    Figure
\ref{figsizevisited} shows the temporal  evolution of the average size
$\langle k_i\rangle$ of the $i^{\rm th}$ visited target, for the $k-l/l_0$
rule and a  superabundant medium with $\beta=2.25$.   For this value
of $\beta$, it  is clear that this quantity  is not stationary.  Three
qualitative  stages  can  be  distinguished.   The  initial  stage  is
characterized  by  the  visit  to  large  targets  all  of  which  are
relatively similar in size; in  a following stage, the size of visited
targets decreases  slowly with time;  in the final stage  (where still
$99\%$ of the targets have  not been visited) the visited targets have
small sizes.

\begin{figure}
\centering      \includegraphics[width=0.35\textwidth]{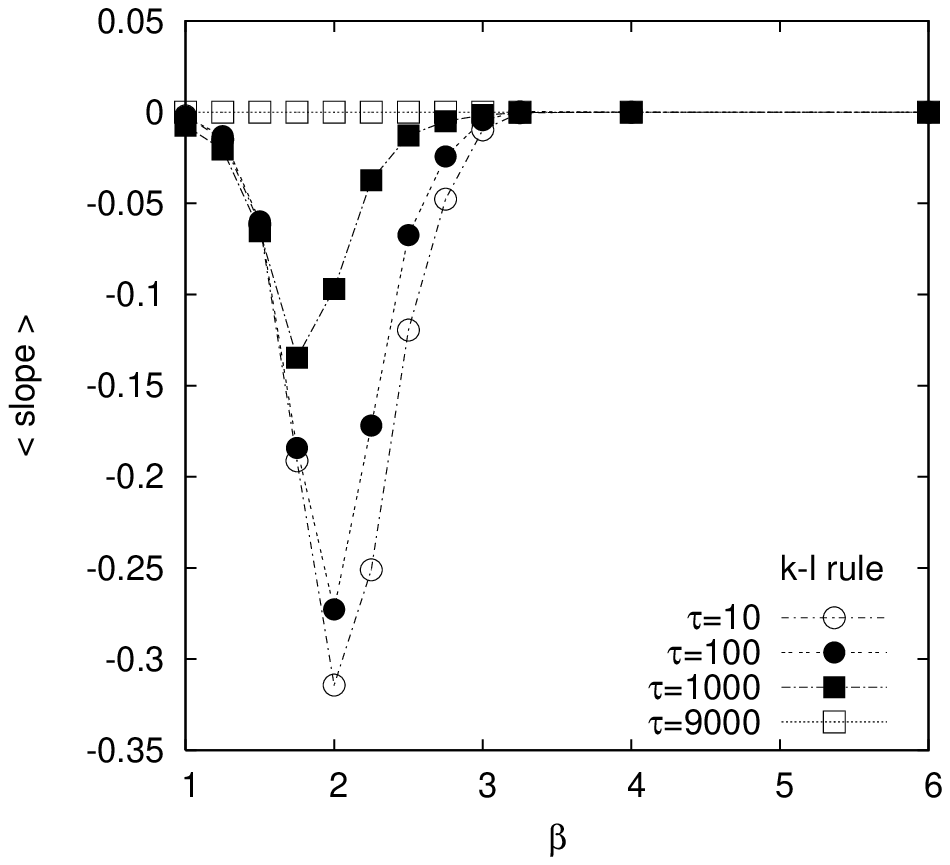}
\includegraphics[width=0.35\textwidth]{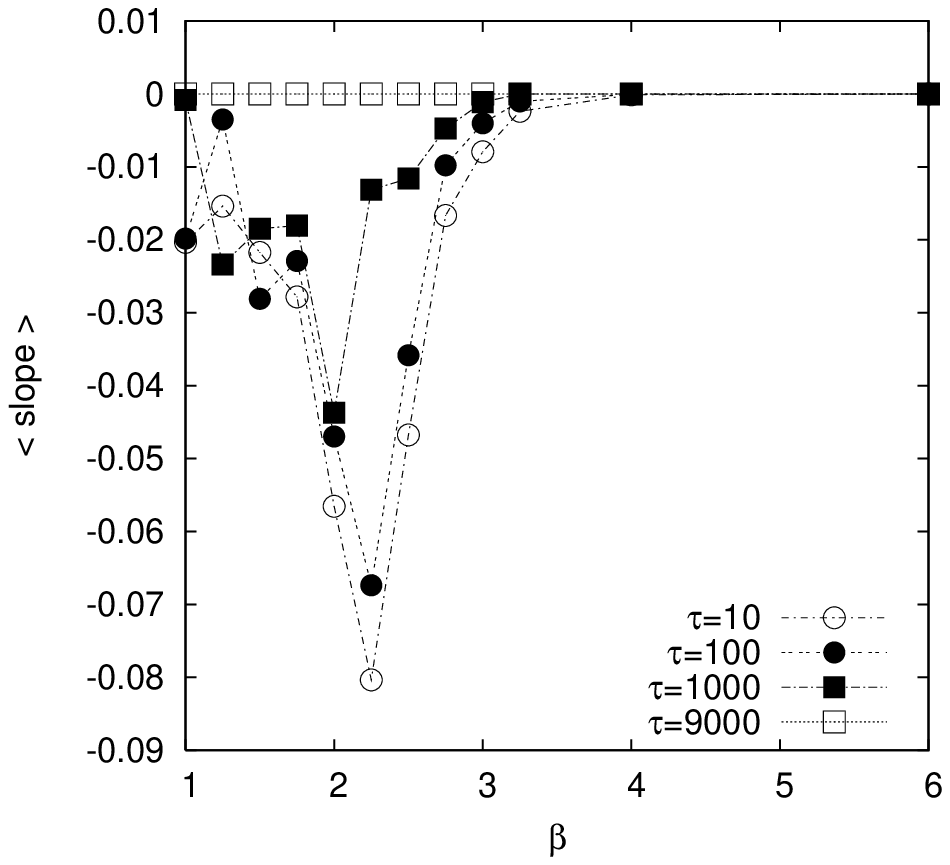}
\caption{Average time derivative
  $\langle\bar{k}_{i+1}-\bar{k}_i\rangle_i$ performed over the time
  domain $i=1,...,10^4$ (for the rules $k-l/l_0$ and $k/l$) as a
  function of $\beta$.  Each point is the average of ten replicas. The
  quantity $\bar{k}_{i}$ is defined as the average value of $\tau$
  consecutive $k_i$ values.  Notice the onset of stationarity at
  $\beta=3$.}
\label{figstationarity}
\end{figure}

This behaviour is typical of the superabundant regime
($\beta<\beta_c$) and can be understood qualitatively in terms of the
length scales considered by the forager. In the initial regime,
decisions are taken essentially in terms of the target size, the
distance travelled playing a minor role (see section \ref{1d} for a
analytical argument). In this regime, large targets are so big that
the walker chooses them in spite of having to take long steps to reach
them, steps that can actually be of the order of the system size. Once
the large targets have been depleted, there are no targets left that
are worth travelling distances of the order of the system size, but
there are still plenty of targets worth travelling distances which are
much longer than the typical inter-target distance $l_0$.  Finally, in
the last regime most of the valuable targets have already been
depleted, and the walker chooses among the many small targets that are
at distances of the order of $l_0$.

Figure \ref{figstationarity} shows the average decay rate of $\langle
k_i\rangle$. It is interesting to note that the process becomes
essentially stationary for $\beta>\beta_c$.  Therefore, in the scaling
regime (\ref{scalnum}), even though the forager modifies its
environment, the medium explored is statistically independent of time
(as long as the number visited target is $\ll N$).

\section{Energy balance and power-laws}\label{secbalance}

Some insights into the properties of the model with the \lq\lq
difference" rule (\ref{kmd}) can be obtained by modifying the energy
cost associated with travels.  We introduce a new parameter, $A$,
which is the energy spent by the forager per unit length $l_0$:
\begin{equation}
E_{ij}=k_j-Al_{ij}/l_0.\label{kmd2}
\end{equation}
In general, we expect that different values of $A$ will induce (in a
same medium) changes in the target choices, that will translate into
changes in the length of the steps and therefore in their
distribution.

\begin{figure}
\hspace{1.5cm} \epsfig{figure=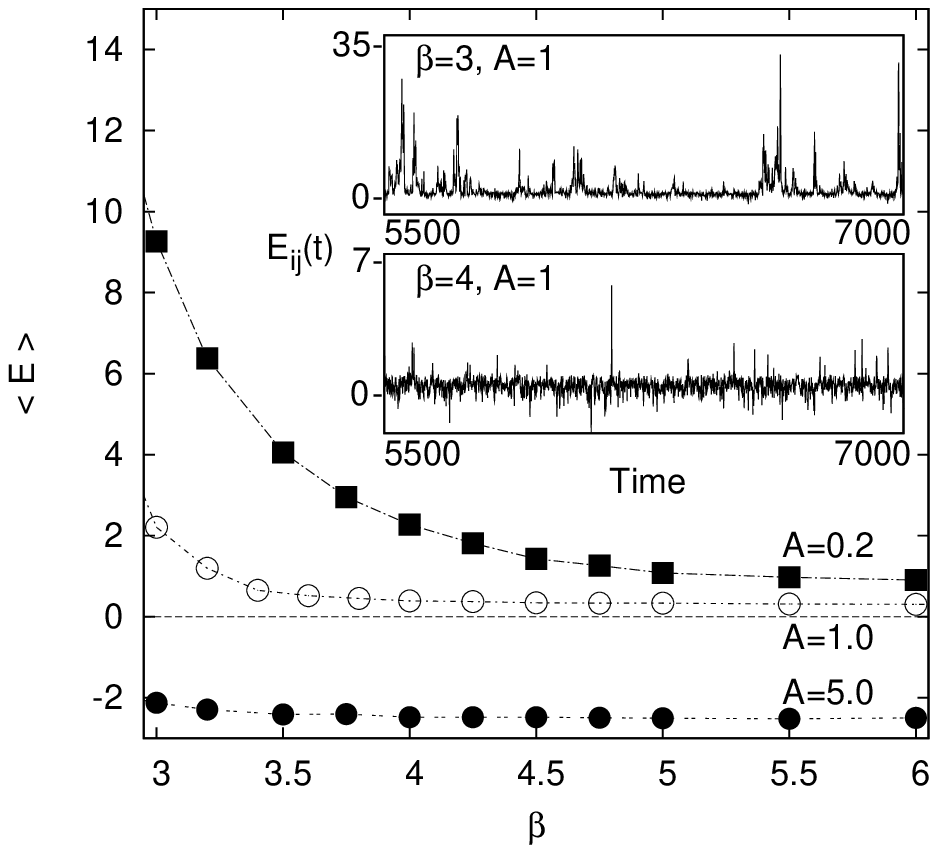,width=3.2in}
\hspace{-1.cm} \epsfig{figure=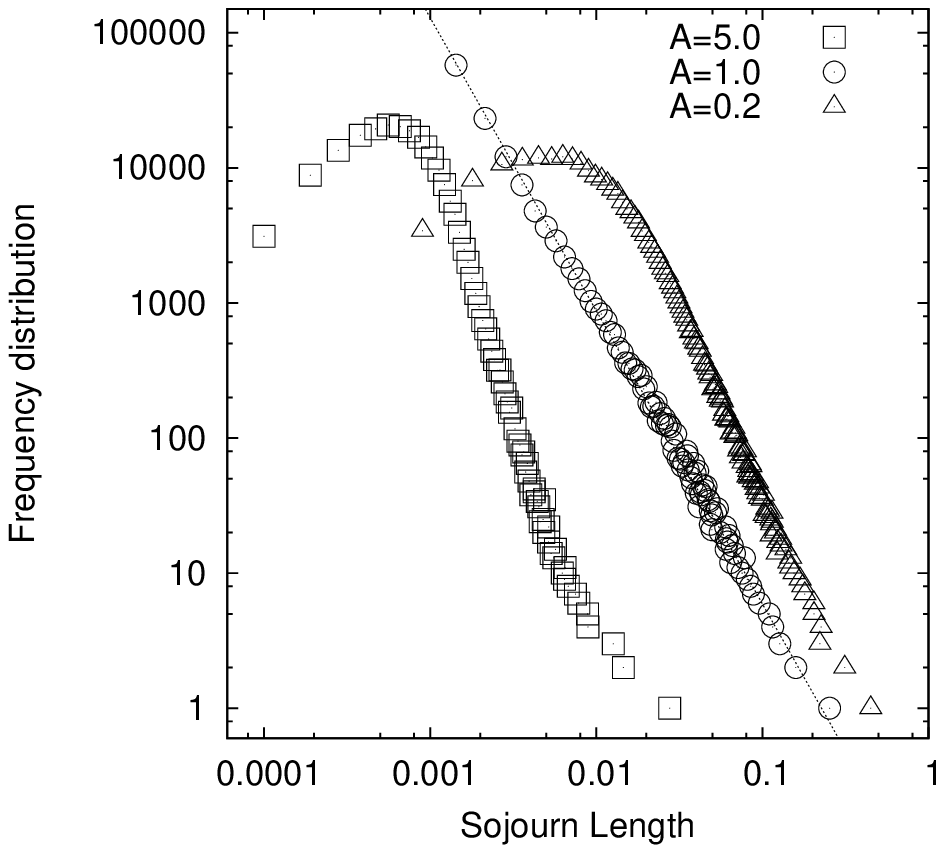,width=2.6in}
\caption{{\bf Left panel:}
Average energy balance as a function of $\beta$ for three values
of $A$ in rule (\ref{kmd2}). Insets: $E_{ij}(t)$ as a function
of $t$ for $\beta=3$ and $4$ ($A=1$ in both cases).
{\bf Right panel:} Step length distribution at fixed $\beta=3$ and
for three values of $A$.}
\label{figbalance}
\end{figure}

The case $A=1$ is the one analyzed so far. If $A\gg 1$, we expect the
average energy balance $\langle E_{ij}\rangle$ to be negative (despite
of the optimal decisions) and the forager starves.  When
$A\ll 1$, on the other hand, we expect the energy gained to
be much larger than the energy consumed in travels. The distances
between successive targets should then not play a major role in the
decisions. Figure \ref{figbalance}, left panel, shows $\langle
E_{ij}\rangle$ as a function of $\beta$, in the stationary regime
({\it i.e.} $\beta\ge 3$), for $A=1,$ $5$ and $0.2$. We observe that
the case $A=1$ is of particular interest and also biologically
relevant, in the sense that $\langle
E_{ij}\rangle >0$ and $\langle E_{ij}\rangle=O(1)$ for all values of
$\beta\ge 3$.  These properties do not hold in the other two
cases. For $A=5$, $\langle E_{ij}\rangle<0$ for all values of
$\beta\ge 3$: a forager with such energy expenditure would not survive
in these media.  For $A=0.2$, we find $\langle E_{ij}\rangle$
is quite larger than 1 at $\beta=3$.

In the insets of the left panel of figure \ref{figbalance}, 
we display typical time series
$E_{ij}(t)$ for $A=1$ and two different values of $\beta$. We observe
in both cases a stationary process.  Remarkably, temporal fluctuations
in the case $\beta=3$ contain a few bursts of size much larger than
the average, a feature reminiscent of a critical phenomena.

The right panel of figure \ref{figbalance} shows the step length 
distribution $P(l)$
at $\beta=3$, for the three values of $A$ considered above. Only the case
$A=1$ clearly exhibits a power-law ($\propto l^{-2}$).
We conclude that power-laws emerge in this model
from the interplay between the energy cost due to travels (that involves 
a space variable, $l_{ij}$) and the energy gain ($k_j$, which is not a space 
variable), when both terms are on average of the same order.

\section{The effect of noise in the choice of a target}\label{1d}

We  analyze  in  this  section  a probabilistic  extension  of  the
deterministic decision rule, based on the \lq\lq quotient"
efficiency (\ref{kod}). 
The approach presents some similarities with that of
\cite{risau1,risau2} where a stochastic version of a deterministic 
walk visiting point-like targets was studied.

In principle, different approaches can be followed to
introduce noise in the model. In one such approach, studied
in \cite{boyer2006}, the forager still seeks to optimism its
movements but has an imperfect knowledge of the medium ({\it e.g},
the values of tree positions and food contents in the mental map
differ from the real ones by some random fluctuating amounts);
hence, the efficiency of a step is not evaluated exactly and the
trajectory can depart from the ideal deterministic one.
For this case, it was shown that in the L\'evy regime these
forager mistakes have little effect on the step-length
distributions \cite{boyer2006}.  Here we focus on a
different approach for introducing noise into the system. In the
present approach the forager still has a perfect knowledge of
the medium and correctly evaluates step efficiencies, but it may
decide not to visit the best target. This mimics the
free-will, so to say, of the agent, allowing for
non-optimal choices.

For simplicity,  let us consider a  one-dimensional lattice  with $N$
regularly  spaced targets  at coordinates  $i=1,2,...,N$ and  a walker
initially located at coordinate  0. Similarly to (\ref{pk}), target
weights are quenched and distributed according to
\begin{equation}
p(k)=(\beta-1)k^{-\beta},\ \beta>1,
\end{equation}
with $1\le k<\infty$ now a continuous variable.  We now
assume that the walker chooses a target $i$ with a probability $p_i$
that depends on the efficiency $E_i=k_i/i$ of the step. We assume
$p_i\propto f(E_i)$, with $f(x)$ an increasing function of $x$.
Hence, high efficiency targets have higher probabilities of being
chosen, but the process is not deterministic. For further simplicity,
we study the $single$ step version of this problem.  A
somewhat similar diffusion problem has been studied in the
weak disorder limit in \cite{belik}. Similar ideas were also
applied to the study of a glass transition in a kinetic trap model
with hopping rates depending on inter-trap distances
\cite{risau1,risau2}.

Let us denote $r_i\equiv f(E_i)$, hence, $p_i=r_i/\sum_j r_j$. In this
section  we  calculate  the  overlap  $Y_2$, defined  as  the  average
probability that two independent walkers  located at the origin in the
same medium choose the $same$ target:
\begin{equation}
Y_2=\langle y_2\rangle,\ {\rm with}\ y_2=\sum_{i=1}^{N}p_i^2,
\end{equation}
where   $\langle.\rangle$   denotes    the   average   over   disorder
configurations  $\{k_i\}$. A  purely deterministic  process
corresponds to $Y_2=1$, whereas a purely random choice leads
to $Y_2\sim 1/N$.

As shown in two examples below,  this problem is similar to the random
energy    model    of    spin   glasses    \cite{derrida}.     Whereas
$Y_2\rightarrow0$   as  $N\rightarrow\infty$   in   the  disorder-free
homogeneous  case ($k_i=constant$), there  is a  $O(1)$, or  even unit
probability  that the  two walkers  choose the  same site  in strongly
heterogeneous media ($\beta<2$, here).

First, we note that
\begin{equation}
y_2=\frac{\sum_{i=1}^{N}r_i^2}{\left(\sum_{i=1}^{N}r_i\right)^2}
=\sum_{i=1}^{N}\int_0^{\infty}dt\   t\  r_i^2e^{-tr_i}e^{-t\sum_{j\neq
i}r_j},
\end{equation}
where  the  identity $A^{-2}=\int_0^{\infty}dt\  t\exp(-tA)/\Gamma(2)$
has been  used. Since the weights are  independently distributed among
targets, then:
\begin{equation}\label{Y21}
Y_2=\sum_{i=1}^{N}\int_0^{\infty}  dt\  t\langle r_i^2e^{-tr_i}\rangle
\prod_{j\neq i}\langle e^{-tr_j} \rangle.
\end{equation}
We  analyze  below two  simple  choices  for  the preference  function
$f(x)$.

\subsection{Exponential preference}

If $f(x)=\exp(x/T)$, where $T$  is a constant, then $r_i=\exp(k_i/iT)$
and  the  deterministic case  is  recovered  as $T\rightarrow0$  (zero
\lq\lq temperature").  At finite $T$, we decompose (\ref{Y21}) into
$Y_2=\int_0^a  dt(..)+\int_a^{\infty}dt(..)   \equiv  Y_2^< +  Y_2^>$,
with $a$  a constant $\ll 1$.  To evaluate the integrand  of the first
integral, we first approximate $\exp(-tr_i)$ to 1 if $r_i<1/t$, and to
0 otherwise. Therefore,
\begin{eqnarray}\label{avexp}
\langle    e^{-tr_i}\rangle&\simeq&\int_1^{-iT\ln   t}p(k)dk=1-(-iT\ln
t)^{-\beta+1} \nonumber\\ &\simeq&\exp[-(-iT\ln t)^{-\beta+1}],
\end{eqnarray} 
an  expansion   valid  for  small   $t$  only.  Given   that  $\langle
r_j^2e^{-tr_j}\rangle=\frac{\partial^2}{\partial      t^2}     \langle
e^{-tr_j}\rangle$,
\begin{equation}\label{av2exp}
\langle r_j^2e^{-tr_j}\rangle=\frac{(\beta-1)(jT)^{-\beta+1}}{t^2(-\ln
t)^{\beta}} \exp[-(-iT\ln t)^{-\beta+1}],
\end{equation}
at     leading      order     as     $t\rightarrow0$.     Substituting
(\ref{avexp})-(\ref{av2exp})  into (\ref{Y21})  and  making the
change $t\rightarrow y=\sum_{j=1}^N (-jT\ln t)^{-\beta+1}$, one obtains:
\begin{equation}\label{Y22}
Y_2^<       =       1-\exp\left[-(-T\ln      a)^{-\beta+1}\sum_{j=1}^N
j^{-\beta+1}\right].
\end{equation}
If $\beta<2$ (superabundant regime in $1d$ \cite{boyer2005}), 
the above sum diverges with $N$:
\begin{equation}\label{Y23}
Y_2^<\simeq                                         1-\exp\left[-(-T\ln
a)^{-\beta+1}\frac{N^{2-\beta}}{2-\beta}\right]   \rightarrow   1\   \
(N\rightarrow\infty).
\end{equation}
Thus, one may conclude that
\begin{equation}
Y_2=1\quad (1<\beta<2),
\end{equation}
like in the deterministic case,  for sufficiently large systems at any
finite  noise  intensity $T$:  The  most  attractive  target at  $i^*$
completely  dominates the  sum  $\sum_{i}r_i$ and  the probability  of
choosing $i^*$  among the  $N$ targets tends  to unity even  at finite
$T$.

On the  other hand, in  the disorder-free case (homogeneous  medium or
$\beta\rightarrow\infty$),
\begin{equation}
Y_2=\frac{\sum_{i=1}^Ne^{\frac{2}{iT}}}
{\left(\sum_{i=1}^Ne^{\frac{1}{iT}}\right)^2}\sim           \frac{1}{N}
\rightarrow0,
\end{equation}
for any  finite $T>0$.  Hence, all targets  become practically equally
attractive and each  walker chooses one of them  essentially at random
(delocalized regime).

\begin{figure}
\centering \includegraphics[width=0.5\textwidth]{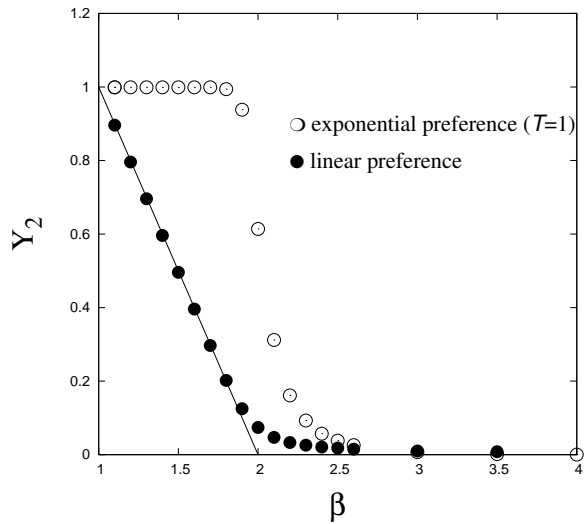}
\caption{Numerically computed probability  $Y_2$ (averaged over $10^4$
disordered media) that two independent walkers choose the same site in
the  one  dimensional problem  with  $N=10^7$.  The  straight line  is
relation (\ref{Y2lfinal}).}
\label{figY2}
\end{figure}

For finite  $\beta>2$, numerical  results indicate that  $Y_2$ remains
finite  as  $N\rightarrow\infty$,  although  it  decays  rapidly  with
$\beta$,  see figure \ref{figY2}.  The values  of $Y_2$  plotted are
practically unchanged  for $N=10^6$  and $N=10^7$.  Hence,  the finite
values  obtained for  $\beta>2$ can  not be  attributed to  finite $N$
effects.

\subsection{Linear preference}

Another case  of interest is the linear  preferential choice $f(x)=x$,
{\it i.e.},  $r_i=k_i/i$. It is exactly soluble  and closely analogous
to the  random energy model \cite{derrida}. In  the disorder-free case
($k_i=1$), $Y_2$ reads
\begin{equation}\label{Y2lindf}
Y_2=\frac{\sum_{i=1}^N i^{-2}}{\left(\sum_{i=1}^N i^{-1}\right)^2}\sim
\frac{\zeta(2)}{(\ln N)^2}\rightarrow0,
\end{equation}
at large $N$.
 
In  the  presence  of   disorder  with  $\beta>2$,  one  has  $\langle
r_i\rangle=j^{-1}(\beta-1)/(\beta-2)<\infty$.   Since  only the  small
$t$  behaviour  contributes in  the  product  in (\ref{Y21}),  then
$\langle    \exp(-tr_i)\rangle\simeq    1-t\langle    r_i\rangle\simeq
\exp(-t\langle r_i\rangle)$ and $\langle r_i^2\exp(-tr_i)\rangle\simeq
\langle  r_i\rangle^2\exp(-t\langle  r_i\rangle)$. Substituting  these
expressions in (\ref{Y21}), one finds:
\begin{equation}
Y_2\sim\int_0^{\infty}du\       u\left(\sum_{i=1}^N      i^{-2}\right)
e^{-\left(\sum_{i=1}^N       i^{-1}\right)u}=       \frac{\sum_{i=1}^N
i^{-2}}{\left(\sum_{i=1}^N i^{-1}\right)^2}\rightarrow0,
\end{equation}
the same result as  the disorder-free case (\ref{Y2lindf}). Therefore,
disorder is irrelevant in heterogeneous media with $\beta>2$.

For  $1<\beta<2$,  $\langle r_i\rangle=\infty$.  It  is convenient  to
write
\begin{equation}\label{interm}
\langle    e^{-tr_i}\rangle=1-(\beta-1)\int_1^{\infty}dk\   k^{-\beta}
[1-\exp(-tk/i)].
\end{equation}
Since $\int_0^1dk\ k^{-\beta}[1-\exp(-tk/i)]$  is integrable and tends
to 0 as $t\rightarrow0$, the lower bound 1 can be replaced by 0 in the
integral of (\ref{interm}). One then obtains,
\begin{eqnarray}\label{interm2}
\langle
e^{-tr_i}\rangle&=&1-(\beta-1)(t/i)^{\beta-1}\int_0^{\infty}dx\
x^{-\beta}                                        (1-e^{-x})\nonumber\\
&\simeq&\exp\left[-(\beta-1)(t/i)^{\beta-1}(-\Gamma(1-\beta))\right],
\end{eqnarray}
for  small  enough $t$.  By  using  (\ref{interm2})  and the  identity
$\langle r_j^2e^{-tr_j}\rangle=\frac{\partial^2}{\partial t^2} \langle
e^{-tr_j}\rangle$,  (\ref{Y21})  can be  integrated  and gives  the
simple result:
\begin{equation}\label{Y2lfinal}
Y_2=2-\beta\quad (1<\beta<2).
\end{equation}
The results are confirmed  numerically in figure \ref{figY2}.  Like in
the  previous exponential  preference, $Y_2$  is non-vanishing  in the
limit  of   large  system  sizes  for   strongly  heterogeneous  media
($\beta<2$). This result is  asymptotically identical to that obtained
for the second moment of the  weights of the microscopic states in the
partition function of the
random energy  model at low  temperatures \cite{derrida}.  It  is also
identical  to  the  $Y_2$  parameter  for a  sum  of  positive  L\'evy
variables \cite{derrida2},  a problem equivalent  to setting $r_i=k_i$
instead of $r_i=k_i/i$ in  (\ref{Y21}). This shows that in strongly
disordered  environments, the  distance travelled  to the  best target
does not play  a major role compared to the target  size in the choice
process.  However, since $Y_2<1$,  noise is not completely irrelevant:
there is still  a finite probability that the  two independent walkers
choose a different target.

\section{Discussion and conclusions}

We  have shown that  a \lq\lq  knowledgeable" walker  following simple
deterministic rules in a medium composed of targets with heterogeneous
weights  describes  trajectories   characterized  by  multiple  length
scales. For  a particular distribution of target  weights, the sojourn
lengths  are maximally  fluctuating and  distributed in  a L\'evy-like
manner, as  $P(l)\sim l^{-\alpha}$ with $\alpha\simeq 2$.  As shown in
\cite{boyer2006},  this model  reproduces  observations of  spider
monkeys foraging  patterns, where $\alpha\simeq 2$  has been measured.
These  monkeys  feed  on  trees  whose  size  distribution  is  broad,
approximately  an  inverse  power-law with  exponent  $\beta\simeq2.6$
\cite{boyer2006}, not far from the special value 
$\beta_c\simeq3$ found in the model.

The type of processes modeled here (but see also \cite{belik} for
a similar approach) also represent a useful alternative to
merely random-walk descriptions of human travels: humans use
(mental) maps, take non-random decisions and concentrate their
activity very heterogeneously in space.  Qualitatively, our model
could describe the movement patterns of hunter-gatherers
\cite{brown} or the diffusion process of bank notes mediated by the
travels of modern humans at the scale of a country
\cite{geisel}. The former system is empirically well described by a
step length distribution with exponent $\alpha\simeq 2$
\cite{brown}, and the latter by $\alpha\simeq 1.6$ \cite{geisel}. In
\cite{geisel}, long enough steps connect different cities and the
power-law behaviour could be induced by the broad distribution of
city sizes, that can be approximated by a Zipf's law \cite{newman},
{\it i.e.}  relation (\ref{pk}) with $\beta=2$. Even-though our
results agree only qualitatively with these observations (no scaling
behaviour in $P(l)$ is actually found here for $\beta=2$), the
present framework, appropriately modified, may provide a
promising alternative for the description of this process. Future
research should consider more sophisticated decision rules and
systems of targets not uniformly distributed in space.

On  a theoretical  point of  view,  the model  proposed exhibits  rich
dynamical features in  spite of its simplicity. We  find, for example,
that  not  every scale-free  weight  distribution  induces walks  with
scale-free  properties.   Besides,  the  fact  that   the  mean  field
approximation worked out in \cite{boyer2005} is not very accurate,
suggests  that memory  effects  (arising the  avoidance of  previously
visited  sites)  play  a  crucial  role. The  special  exponent  value
$\alpha=2$ found in simulations is actually not well understood.

We also  present arguments showing that  in the presence  of strong
disorder (small values of $\beta$ here), the addition of stochasticity
in  the choice  rules should  not affect  the main  properties  of the
walks,  provided that  the probability  of visiting  a  site increases
sufficiently rapidly with the  site attractiveness.  This situation is
similar  to  the  lack  of  self-averaging  in  spin  glasses  at  low
temperature \cite{derrida2}. As shown  by our calculation of $Y_2$ in
the   one-dimensional  case,   noise  should   not  affect   much  the
distribution  of sojourn  length $P(l)$,  thus explaining  
{\it a posteriori} the results found in \cite{boyer2006} in a  variant 
of the model including forager mistakes. 
The analytical  results presented  here being limited  to  a single  step
process, it would be  interesting to further investigate whether there
is still a finite probability to find two (or $n$) independent walkers
at  the same  site  after  many steps.  Such  probability does  remain
asymptotically  finite in  several  random walk  models in  disordered
media, the remarkable  Golosov  localization phenomenon \cite{monthus}.

A system of many non-interacting travelling agents with mental map are
susceptible to make contact (when they occupy a same site) much more
frequently than random walkers.  The results above suggest that
such contact networks could be, in some cases, controlled by the
quenched randomness of the medium in spite of the
stochasticity of the agent's decisions.  Such robustness could
have important social implications, for instance for the formation of
social networks or the spreading of diseases \cite{boyer2006b}.

\ack

D.B.  and  O.M.  acknowledge  support from DGAPA-UNAM  IN-118306;
H.L.  acknowledges  support from DGAPA-UNAM  IN-112307.  The
authors thank  G. Ramos-Fern\'andez, J.L.  Mateos, G.  Cocho, H. Ramos
and F. Rojas for previous collaboration on this problem.

\end{document}